\documentclass[a4paper]{article}
\usepackage{latexsym}
\usepackage{cite}
\usepackage{epsfig,multicol}
\usepackage{amsmath,amssymb,mathrsfs}
\usepackage{fullpage}

\newcommand\lsim{\mathrel{\raise.3ex\hbox{$<$\kern-.75em\lower1ex\hbox{$\sim$}}}}
\newcommand\gsim{\mathrel{\raise.3ex\hbox{$>$\kern-.75em\lower1ex\hbox{$\sim$}}}}

\newenvironment{Eqnarray}%
     {\arraycolsep 0.14em\begin{eqnarray}}{\end{eqnarray}}
\newcommand{\beqa}{\begin{Eqnarray}}
\newcommand{\eeqa}{\end{Eqnarray}}
\newcommand{\beq}{\begin{equation}}
\newcommand{\eeq}{\end{equation}}
\begin{document}

\title{
A 2D fit with background measurement constraints 
to boost the Higgs~$\rightarrow$ZZ$^{(*)}\rightarrow$4$\ell$ 
discovery potential at the LHC }
\author{Christos Anastopoulos$^{1}$, Nicolas Kerschen$^{1}$ and Stathes Paganis$^2$\\ 
$^{1}$ Department of Physics, CERN, CH-1211, Geneve 23, Switzerland\\
$^{2}$ Department of Physics and Astronomy, University of Sheffield, S3 7RH, UK\\
}
\maketitle
\begin{abstract}
A data-driven method for simultaneously extracting 
a potential Higgs$\rightarrow$ ZZ$^{(*)}\rightarrow$4e, 4$\mu$, 2e2$\mu$ 
signal and its dominant backgrounds, is presented.
The method relies on a combined fit of 
the 2-lepton, Z$^{(*)}$, and 4-lepton invariant masses.
The fit is assisted by normalization of the Z+X backgrounds in data control 
regions. The Higgs discovery potential for the next few years of LHC running 
is presented. The demonstrated high sensitivity of the method 
makes it ideal for the search performed by the ATLAS and CMS experiments.
\end{abstract}

\section{Introduction}
\label{intro}
The discovery of the Standard Model (SM) Higgs boson is the major goal 
of the Large Hadron Collider (LHC). With the 2011 proton-proton 
experimental run now underway, LHC experiments are entering the 
main phase of the Higgs search. The Higgs boson 
mass is a free parameter in the SM, however there is strong
expectation motivated by precision electroweak data \cite{EWeak} 
and direct searches \cite{Higgs}, that a
low mass Higgs ($114.4-186$~GeV\footnote{Natural units are used throughout this paper.}, at 95\% confidence level) 
should be discovered at the LHC.

The experimentally cleanest signature for the discovery of the 
Higgs is its ``golden'' decay to four leptons (electrons and muons): 
H$\rightarrow$ZZ$^{(*)}$$\rightarrow$4$\ell$.
The excellent energy resolution and linearity of the reconstructed
electrons and muons leads to a narrow 4-lepton invariant mass peak 
on top of a smooth background. The expected signal to background 
ratio after all experimental requirements ranges approximately from 1/1 to 5/1 
depending on the Higgs mass \cite{CSC}.
The major component of the background consists of the irreducible 
pp$\rightarrow$ZZ$^{(*)}$$\rightarrow$4$\ell$ decays. 
The most challenging mass region is between 120-180~GeV where one 
of the Z bosons is off-shell giving low transverse momentum 
leptons. In this region backgrounds from 
pp$\rightarrow$Zb$\bar{\mathrm{b}}$, Zc$\bar{\mathrm{c}}\rightarrow$4$\ell$ (denoted as ZQQ) 
and to a lesser extend Zjj and ZjQ (j denotes a light quark which is mis-identified
as a lepton)  
are significant, requiring tight lepton isolation cuts in order to reduce 
these backgrounds to levels well below the pp$\rightarrow$ZZ$^{(*)}$ background.

In order to 
maximize the Higgs discovery potential using the H$\rightarrow$4$\ell$
channel, 
the most accurate achievable knowledge of the background and its 
associated uncertainty are essential. The main dominant 
pp$\rightarrow$ZZ$^{(*)}\rightarrow$4$\ell$ 
background can be estimated by (i) theoretical predictions, 
(ii) by a combination of theoretical predictions and subsequent 
constraints using experimental LHC data, or (iii) by performing 
a sideband measurement and subsequently extrapolating to the signal
region (SR). 
As discussed in \cite{CMSTDR}, the first two methods suffer from 
theoretical uncertainties, the luminosity measurement uncertainty 
(only for method (i))  and systematic uncertainties
such as lepton reconstruction efficiency.
In addition, for low Higgs mass, the ZQQ contribution is 
significant with a theoretical uncertainty 
ranging from $20-50\%$ \cite{campbell} 
(ideally it should be experimentally controlled 
with early LHC data).

Data-driven background extractions using
the sideband measurement (iii) have been considered by both ATLAS 
and CMS \cite{CSC}, \cite{CMSTDR}.
Such extractions are not as sensitive to theoretical
and luminosity uncertainties, and uncertainties due to lepton and 
isolation efficiencies, but for relatively low integrated luminosity 
are limited by the number of events in the  4-lepton sideband \cite{Paganis}.
Thus, a method providing a simultaneous 
measurement of a potential Higgs signal and all its 
background components is essential for the Higgs discovery
in the most important LHC channel.
This method should be able 
to give a realistic recipe for the extraction of all major components 
of the background (Zjj, ZQj, ZQQ, ZZ$^{(*)}$) while improving 
the power of the Higgs search.
In this work we propose such a novel method 
applicable in early LHC measurements. 

Our main result is that 
using direct measurements of all backgrounds and a combined fit 
of the 2-lepton and 4-lepton invariant masses, the Higgs should 
be discovered with 2-5 $fb^{-1}$ of integrated luminosity  
combining the analyses of the two LHC experiments.
The proposed method can be directly employed in the Higgs searches 
currently performed by the ATLAS and CMS experiments \cite{ATLASEPS2011}, \cite{CMSEPS2011}.

\section{Description of the Method}

The main goal of the technique presented in this paper is to 
optimize the Higgs discovery potential using all 
available information from data
by simultaneously extracting the 
Z+X$\rightarrow$4$\ell$ and ZZ$^{(*)}\rightarrow$4$\ell$ 
yields which are expected to fully dominate the $4\ell$ continuum.
This is achieved by a 
synthesis of a 2D fit assisted by constraints in the 
normalizations of the Z+X  backgrounds using data control regions.
The Z+X reducible background consists of 
three components Zjj (Z+2jets), ZjQ (Z+1jet plus a lepton from 
a heavy flavour decay) and ZQQ. The jet contribution includes
photons. 
The method will be applied for a low mass Higgs search (M$_H<2\times M_Z$)
and a high mass search (M$_H\geq 2\times M_Z$).

In the study presented here the proton-proton collisions at a center 
of mass energy of 7 TeV were generated 
using PYTHIA \cite{PYTHIA}. A fast generic detector simulation was
performed by smearing the generated values in order 
to include acceptance, resolution and efficiency effects.
The ZQQ and ZQ fractions of Z+X were obtained using MCFM \cite{campbell}.

\subsection{Event selection}

A typical event selection for the H$\rightarrow$4$\ell$ analysis is 
as follows:
\begin{itemize}
\item
Select events with at least 4 leptons with transverse momentum $(p_{\perp})>$7~GeV in pairs of same flavour 
and opposite charge. At least one of these leptons must have high $p_{\perp}>$20~GeV.
\item
Require one on-shell Z$\rightarrow$2$\ell$ boson. The
on-shell requirement is satisfied by an invariant mass (M$_{12}$) 
cut of $\pm$ 15~GeV about the nominal Z-mass. 
The 4-lepton requirement leads to an on-shell Z invariant mass 
distribution which has a small flat component of reducible backgrounds.
\item
The two leptons involved in the on-shell Z reconstruction 
are required to have a strict level of isolation. This
is achieved using the tracker and the calorimeter. They are also 
required to originate from the interaction vertex.
\item
The remaining 2 same-flavour leptons have an invariant mass
M$_{34}$ which is the key discriminating variable 
in this analysis. It will also serve as the ZZ data control region. 
The cuts on these 2 leptons depend on the mass hypothesis 
(high or low Higgs mass). 
\end{itemize}
Non-Z reducible backgrounds ($\ensuremath{\mathrm{t}\bar{\mathrm{t}}}$ and $\ensuremath{\mathrm{W}}+$jets) 
after event selection are very small and can be subtracted 
by removing the smooth background under the Z mass peak. 
Leptons from light quarks are non-isolated, while those 
from heavy quarks (c,b) have in addition a 
significant impact parameter (normalized distance of closest approach) 
from the interaction vertex. 

In this work a selection using stringent
isolation and impact parameter criteria is applied for the
low Higgs mass search, whereas for the high Higgs mass
search no selection is applied on the impact parameter and
the isolation criteria are relaxed. Typical signal and background
yields for the two event selections performed are
shown in Table~\ref{xsections}. The relaxed selection criteria for the
high Higgs mass search are motivated by the fact that
the Z+X background does not peak in high values of the
4-lepton mass, M$_{4l}$. This background can be estimated by the 2D fit using
the information in M$_{34}$.
\begin{table}
\small
  \begin{center}    
    \begin{tabular}{|l|c|c|}\hline
      Expected &M$_H$=150GeV (fb)&M$_H$=240GeV (fb)\\\hline
      Higgs & 2.19  & 3.68 \\\hline
      ZZ$^{(*)}$ & 11.05 & 15.14 \\\hline
      Zjj & 0.0009 & 0.04 \\\hline
      ZQj & 0.025 & 0.57 \\\hline
      ZQQ & 0.56  & 6.13 \\\hline   
      Z+X & 0.582 & 6.74 \\\hline 
    \end{tabular}
    \caption{ Expected 4-lepton event yields (in the M$_{4l}$ mass
      range) per 1 fb$^{-1}$ of integrated luminosity 
      after a low mass analysis selection
      (first column), and a high mass analysis selection (second column) 
      for proton-proton collisions at a center of mass
      energy of 7 TeV. Detector effects such as acceptance,
      efficiencies and resolutions have been applied via a fast
      simulation (see text).} 
    \label{xsections}
  \end{center}
\end{table}

The background components can be constrained using measurements in 
data control regions defined by a subset of the analysis cuts.
Each of these data control regions is rich in a particular 
background component. 
The definition of such data control regions  
(Zjj, ZQj) involves the two additional subleading leptons. 
Since lepton isolation cuts 
lead to a large rejection of the Zjj and ZjQ components,
it is a common experimental practice to reverse such cuts 
in order to define data control regions rich in these backgrounds
\cite{ATLASEPS2011}.
An example is given in the next section.
\subsection{Lepton Isolation: a data control region example}
\label{seclepiso}
In this section we 
present an example of a usage of a data control region to predict the
reducible background in the signal region.

\begin{figure}[ht]
\centering
\resizebox{0.5\textwidth}{!}{%
\includegraphics{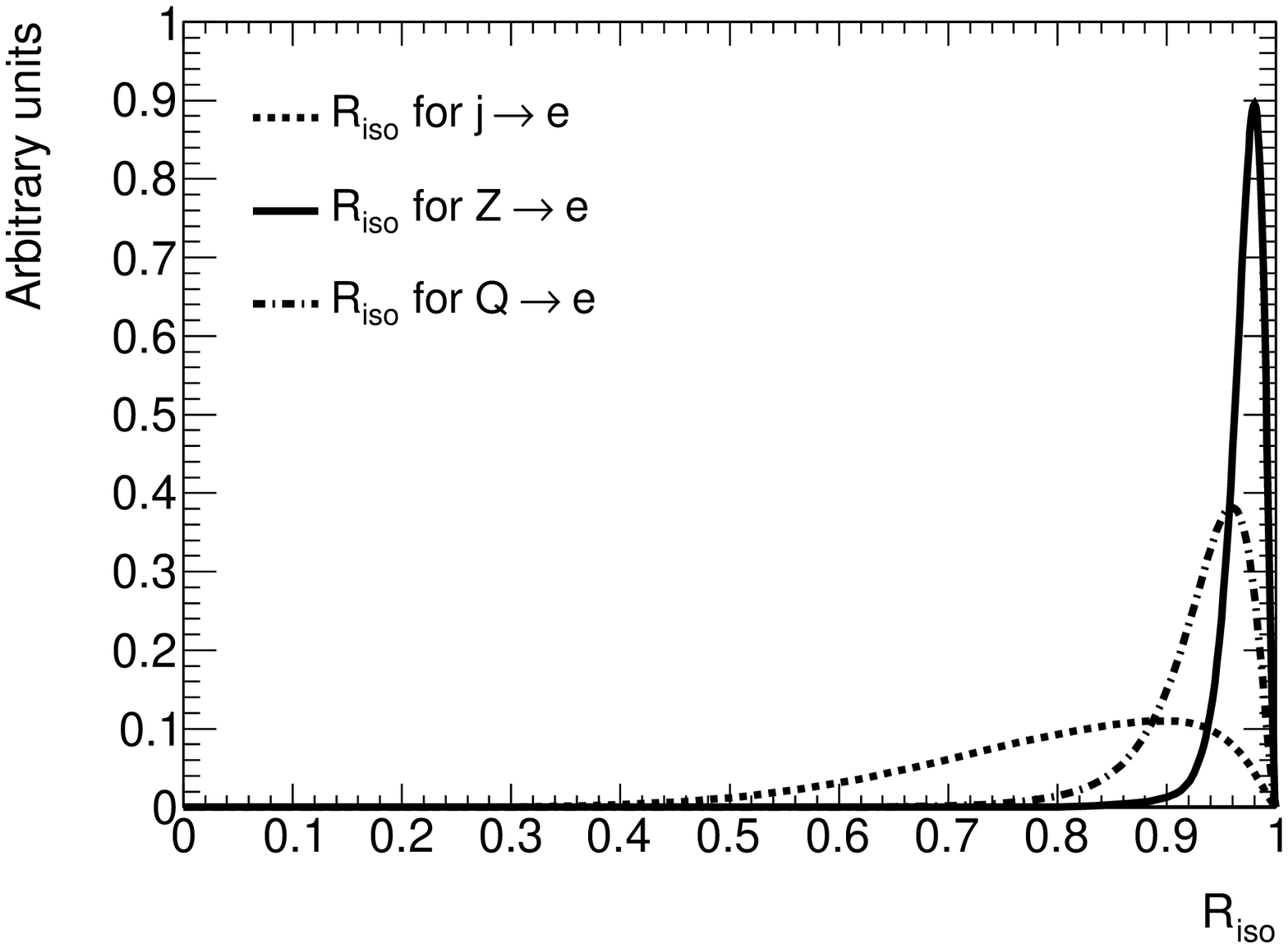}
}
\resizebox{0.5\textwidth}{!}{%
\centering
\includegraphics{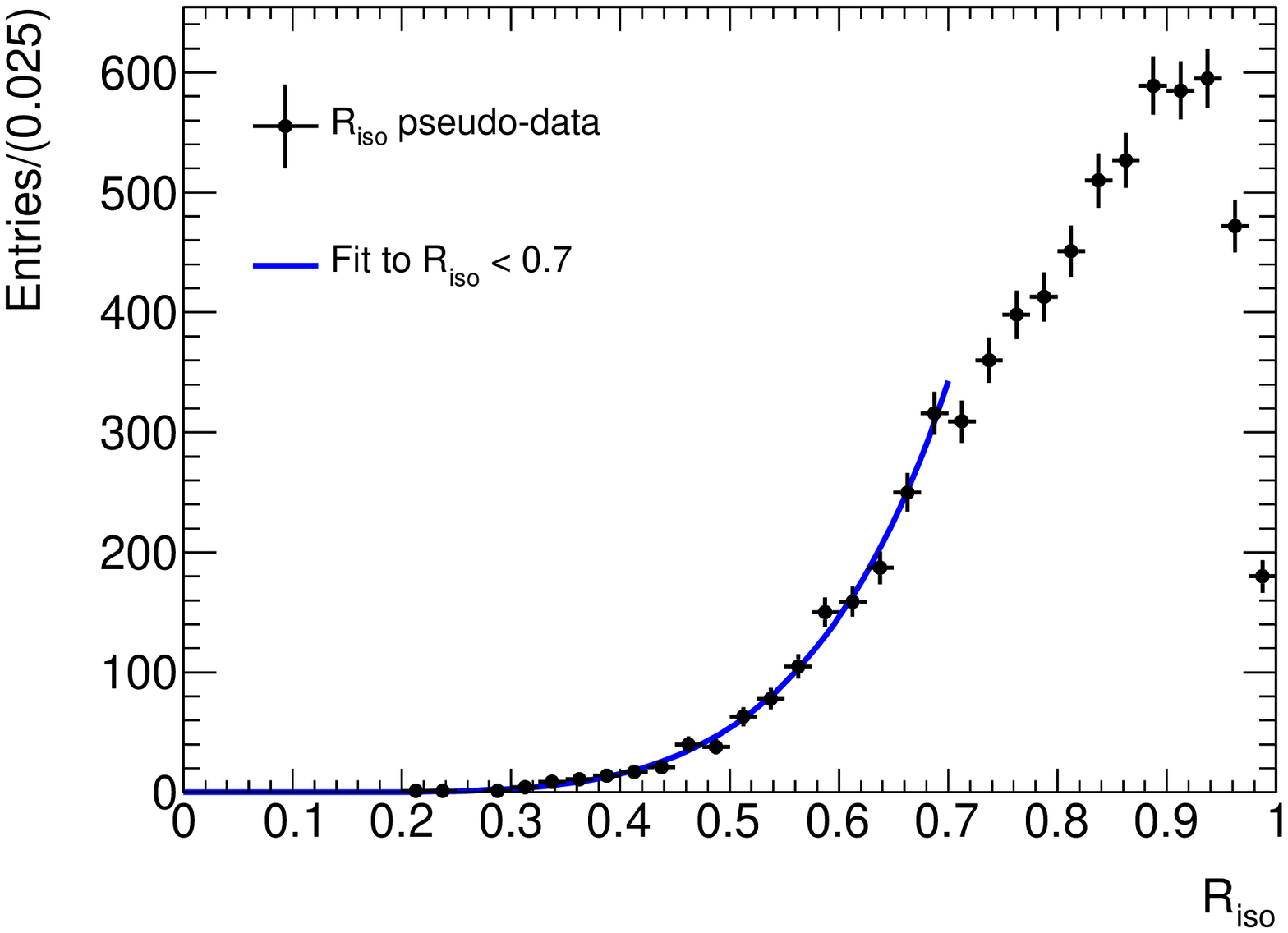}
}
\caption{
Top: distributions of containment variable $R_{iso}$ for isolated
electrons (continuous line), electrons from heavy flavour decays (dashed-dotted line) and electrons 
originating from light quark jets (dashed line). Bottom: 
Isolation $R_{iso}$ distribution for a random pseudo-experiment.
For $R_{iso}<0.7$, the fit used for the normalization of the 
Zjj background to the data is overlayed.}
\label{Reta}
\end{figure}

To demonstrate how these regions can be used for electrons to extract
the normalizations of 
the Zjj and ZjQ backgrounds after all analysis cuts, we use 
an electromagnetic (EM) shower containment variable $R_{iso}$~\cite{ATLASEPS2011}.
We define $R_{iso}=E_{in}/E_{tot}$ as the ratio of EM energy 
inside a 0.075$\times$0.125 cone, divided by the energy 
in a 0.125$\times$0.125 cone. The distributions of $R_{iso}$ for isolated 
electrons from Z-bosons and for non-isolated leptons are shown 
in Figure~\ref{Reta}. Its shape was obtained from existing measurements 
at the LHC~\cite{eperf}.
For $R_{iso}<0.7$ and given the cross sections of 
Table~\ref{xsections}, $R_{iso}$ is fully dominated by 
j$\rightarrow$e fakes. This $R_{iso}<0.7$ region provides a data control region
for Zjj. The normalization of the Zjj background after 
all analysis cuts can be extracted if the data events found in the 
data control region with $R_{iso}<0.7$ are normalized to a Zjj Monte Carlo
(Figure~\ref{Reta} bottom),
and subsequently the remaining analysis cuts are applied to the MC.
This final step (the extrapolation with MC) relies on a good 
control of the additional discriminants used \cite{ATLASEPS2011}. 
Typical such variables used by ATLAS and CMS are: 
isolation, EM shower shape variables and impact parameter.
The predicted number of Zjj background events in the signal region is 
given by the following expression:
\begin{equation}
  N_{Zjj}(\mathrm{Signal~Region}) = N^{CR}_{Zjj} \times 
  \frac{N^{Data}_{R<0.7}}{N^{MC}_{R<0.7}} \times  
  \epsilon_{MC}, \nonumber
  \label{eq:zjj}
\end{equation}
where $N^{CR}_{Zjj}$ is the number of Zjj events in the Zjj control
region in MC and $\epsilon_{MC}$ is the efficiency for Zjj events
passing the final cut selection. 

\subsection{Extraction of background contributions}
In this work we introduce a novel technique in which the 
signal and all background components are fitted simultaneously in the 
M$_{34}$ and M$_{4\ell}$ invariant mass observables.
This is a 2D unbinned extended likelihood fit which exploits 
constraints coming from independent measurements of the Z+X backgrounds in 
data control regions. The method is applied on two 
separate searches:
a low mass search ($M_H<2\times M_Z$) 
where one Z is off-shell, and a 
high mass search ($M_H\geq 2\times M_Z$) 
where both Z's can be on-shell.
The two regions are dominated by different background 
fractions and shapes, thus a separate treatment is 
required.

\begin{figure}[htb]
\centering
\resizebox{0.5\textwidth}{!}{%
\includegraphics{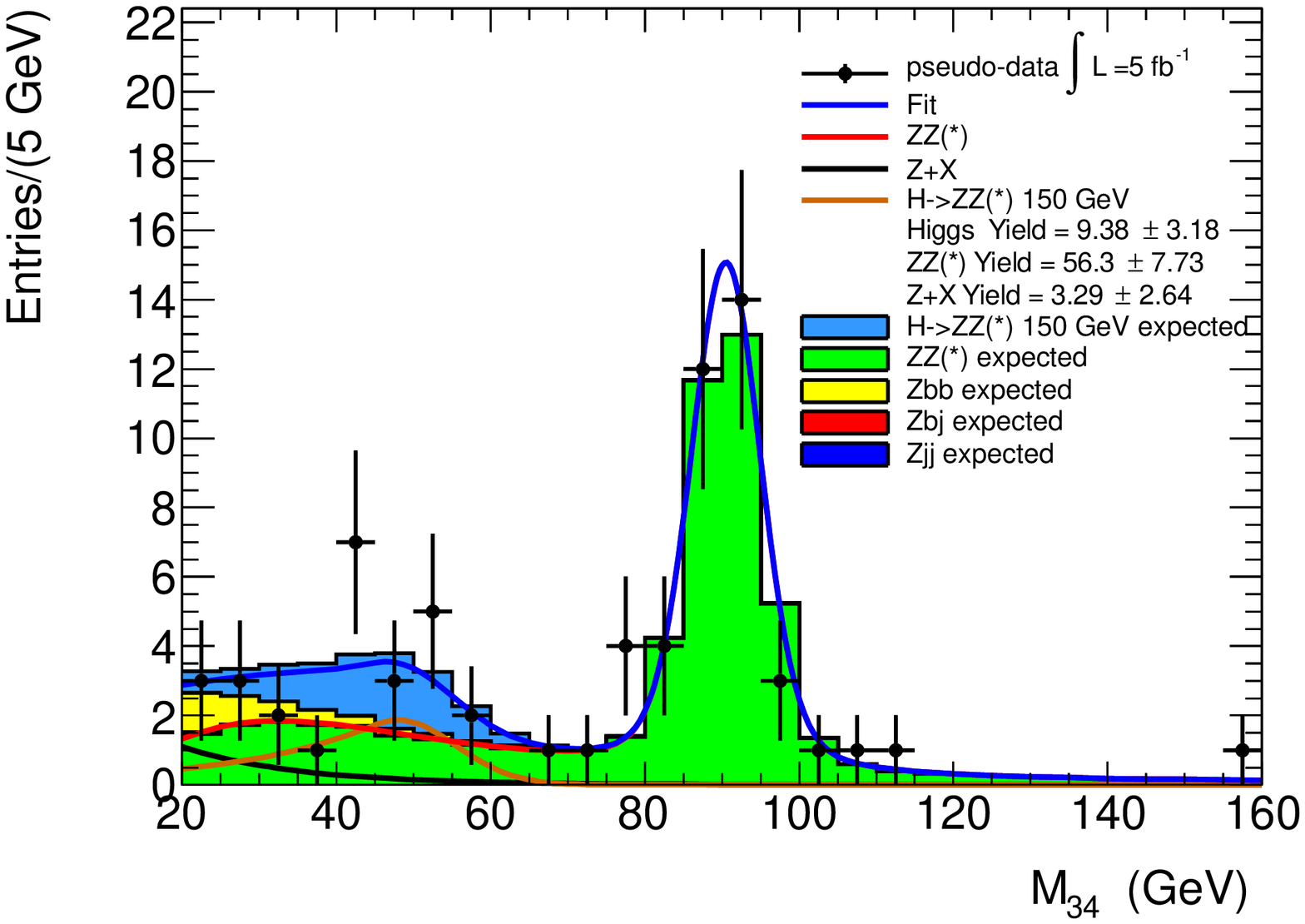}
}
\centering
\resizebox{0.5\textwidth}{!}{%
\includegraphics{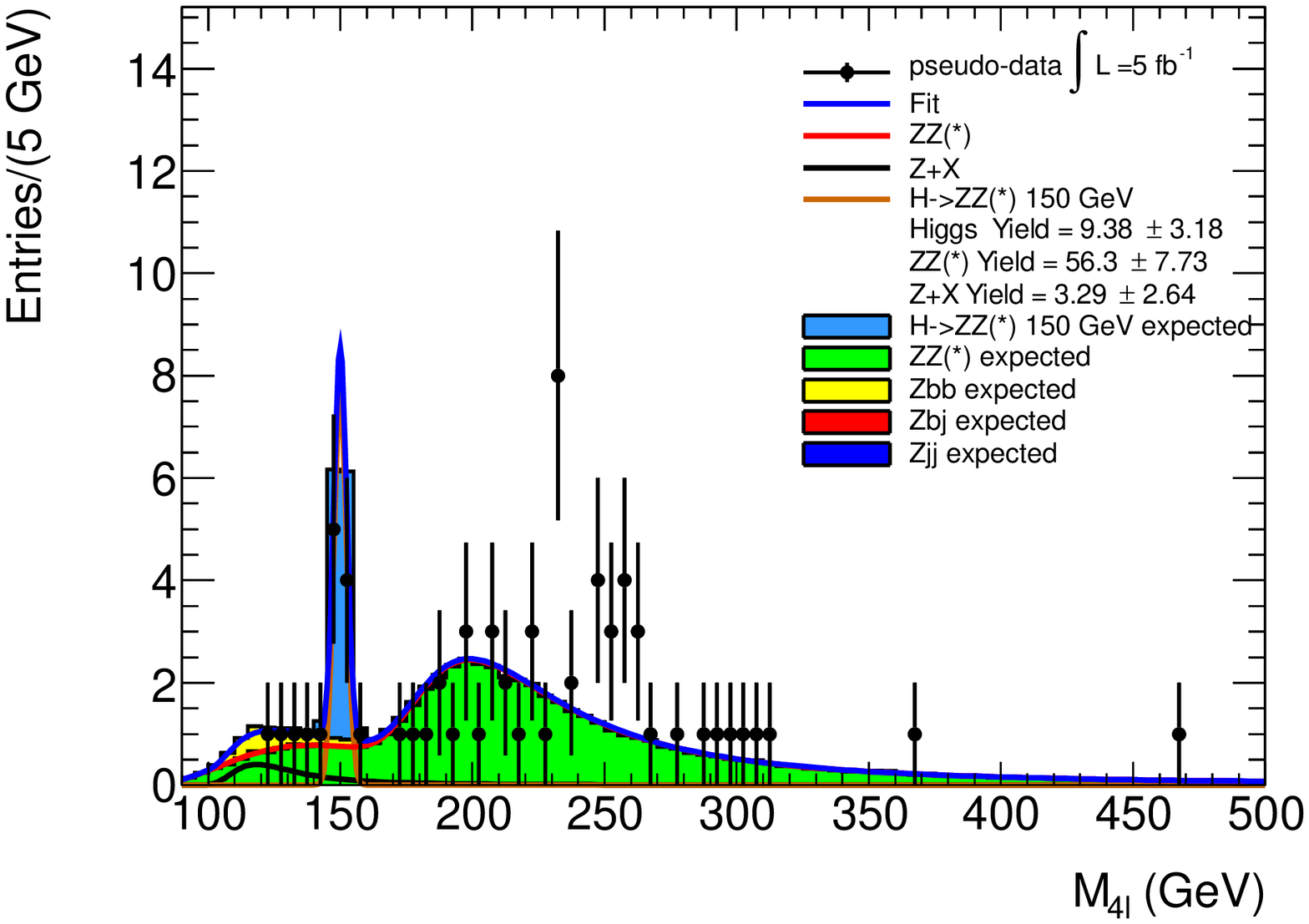}
}
\caption{
Simultaneous unbinned extended likelihood fit of
the 2-lepton M$_{34}$ and 4-lepton M$_{4l}$ 
invariant masses for a low mass Higgs ($M_H<2\times M_Z$) and 
an integrated LHC luminosity of 5 fb$^{-1}$. The fit exploits 
measurements of the Z+X backgrounds in data control regions to 
extract both the signal and backgrounds after the final selection
(see text).}
\label{lowFit}
\end{figure}

\begin{figure}[htb]
\centering
\resizebox{0.5\textwidth}{!}{%
\includegraphics{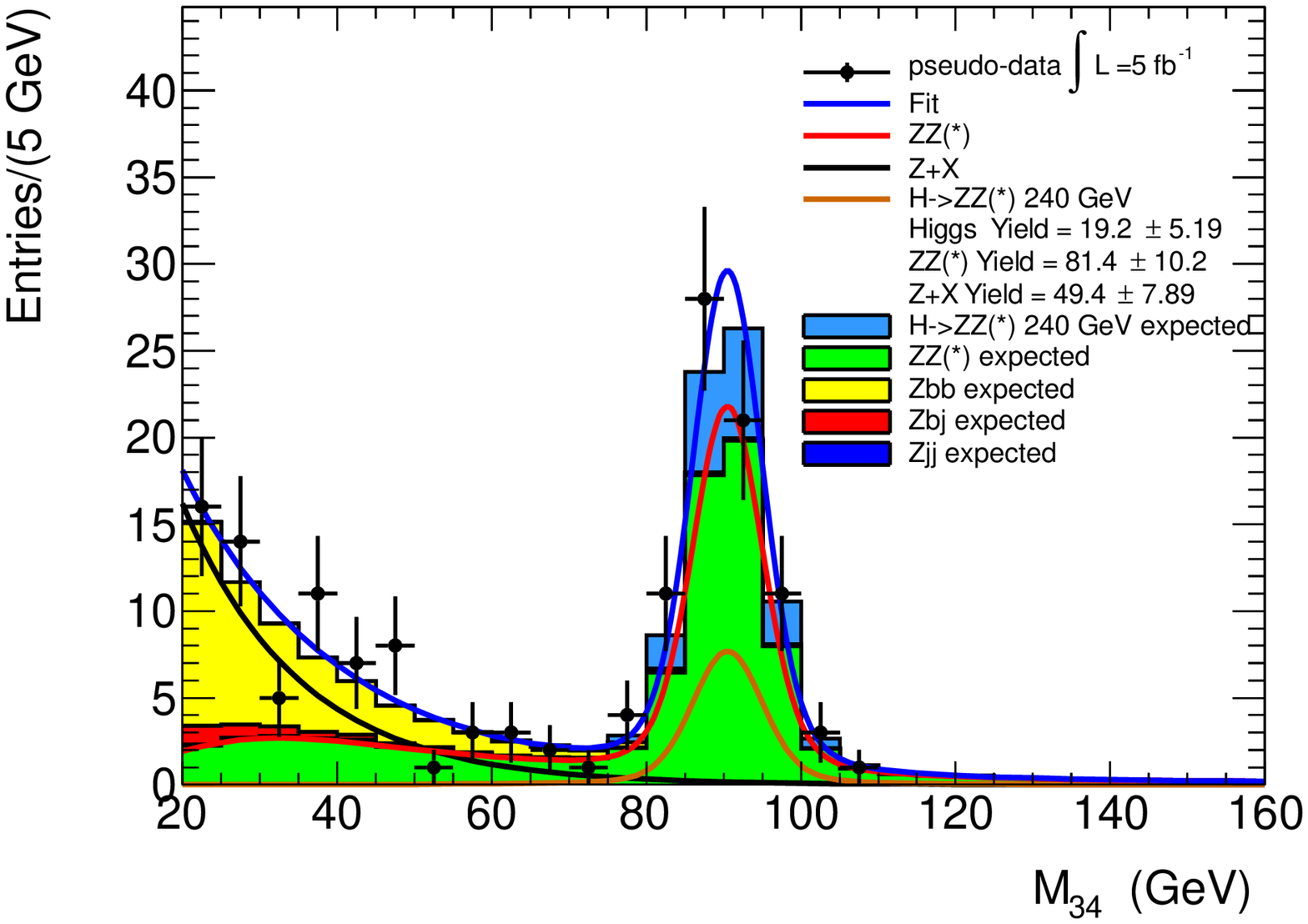}
}
\centering
\resizebox{0.5\textwidth}{!}{%
\includegraphics{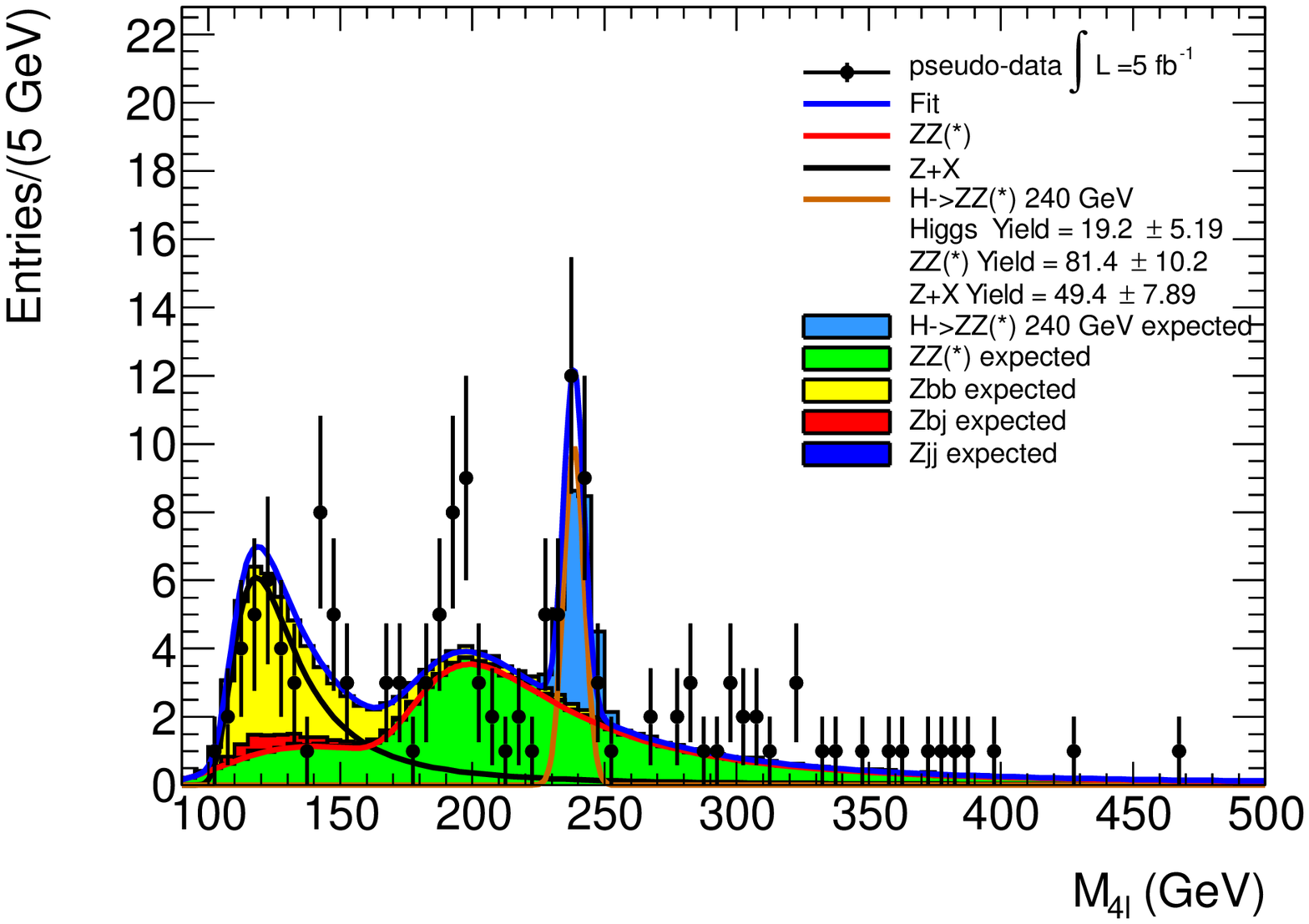}
}
\caption{
Simultaneous unbinned extended likelihood fit of 
the 2-lepton M$_{34}$ and 4-lepton M$_{4l}$ 
invariant masses for a high mass Higgs ($M_H\geq 2\times M_Z$) and 
an integrated LHC luminosity of 5 fb$^{-1}$. The fit exploits 
measurements of the Z+X backgrounds in data control regions to 
extract both the signal and backgrounds after the final selection
(see text).}
\label{highFit}
\end{figure}

Figure~\ref{lowFit} shows a simultaneous fit 
of the low Higgs mass M$_{34}$ and M$_{4\ell}$ distributions 
for pseudo-data corresponding to an integrated 
luminosity of 5 fb$^{-1}$ for a single LHC experiment. 
Figure~\ref{highFit} shows similar fits for the high 
mass search. 
The method exploits the fact that the Higgs distribution makes 
a narrow peak in the M$_{4\ell}$ distribution, while the backgrounds 
are smoothly distributed. In the high mass region, the ZZ signal 
is clearly visible in M$_{34}$, however, it could be contaminated by
a potential Higgs signal. These two components can be easily separated using 
the information in the M$_{4\ell}$ distribution (Figure~\ref{highFit} bottom).
The Z+X component for the high Higgs
mass search is significant, due to the relaxed selection used
in this work, but it can be easily estimated using the information
in the M$_{34}$ region (Figure~\ref{highFit} top). 
These example 2D fits are performed using the normalization constraints obtained 
from data summarized below:
\begin{itemize}
\item
The excellent separation between ZZ and the rest of the 
backgrounds  in the 
M$_{34}$ distribution provides the fit with a powerful 
discriminant. This allows the extraction of normalizations 
for the ZZ and the 
rest of the backgrounds.
In this study the ZZ and Z+X components are
fitted with a Breit-Wigner convoluted with a gaussian resolution function
and an exponential respectively.
\item
At low values of M$_{34}$ the ZZ* component is obtained by the ZZ 
normalization from data and the relative ZZ*/ZZ normalization 
from MC. The residual uncertainty on the ZZ* normalization is 
taken as a systematic, since there is no other experimental 
input to constrain ZZ*. 
\item
The normalizations of the ZQj and Zjj backgrounds 
can be constrained by measurements in the control regions. 
These normalizations have uncertainties 
coming from the experimental knowledge of the 
relevant discriminants used in the extrapolation 
from the control region to the signal region (see section~\ref{seclepiso}). 
\item
Finally the normalization of ZQQ is let to float in the fit
since it is fully unconstrained. Constraints of ZQQ are possible
by requiring large impact parameters for the leptons, offering the
possibility to directly measure the ZQQ component in Z+X. This is
beyond the scope of this work.
\end{itemize}

As shown in Figures~\ref{lowFit} and~\ref{highFit}, the potential 
Higgs signal and its associated backgrounds can be extracted in a 
single fit assisted by constraints on the components using data.

\section{Results: Higgs discovery potential vs Luminosity}

The 2D fit described in the previous section 
leads not only to an extraction of the signal and all of 
the major backgrounds (ZZ, ZQQ, ZQj, Zjj$\rightarrow$4$\ell$), 
but also to the signal significance~\cite{pdg}.
In this section we use the method to assess the SM Higgs discovery 
potential. For the extraction of the significance the 
RooStats package of the RooFit/ROOT analysis framework has been used
\cite{RooStats}, \cite{roofit}, \cite{ROOT}.

\begin{figure}[htb]
\centering
\resizebox{0.5\textwidth}{!}{%
\includegraphics{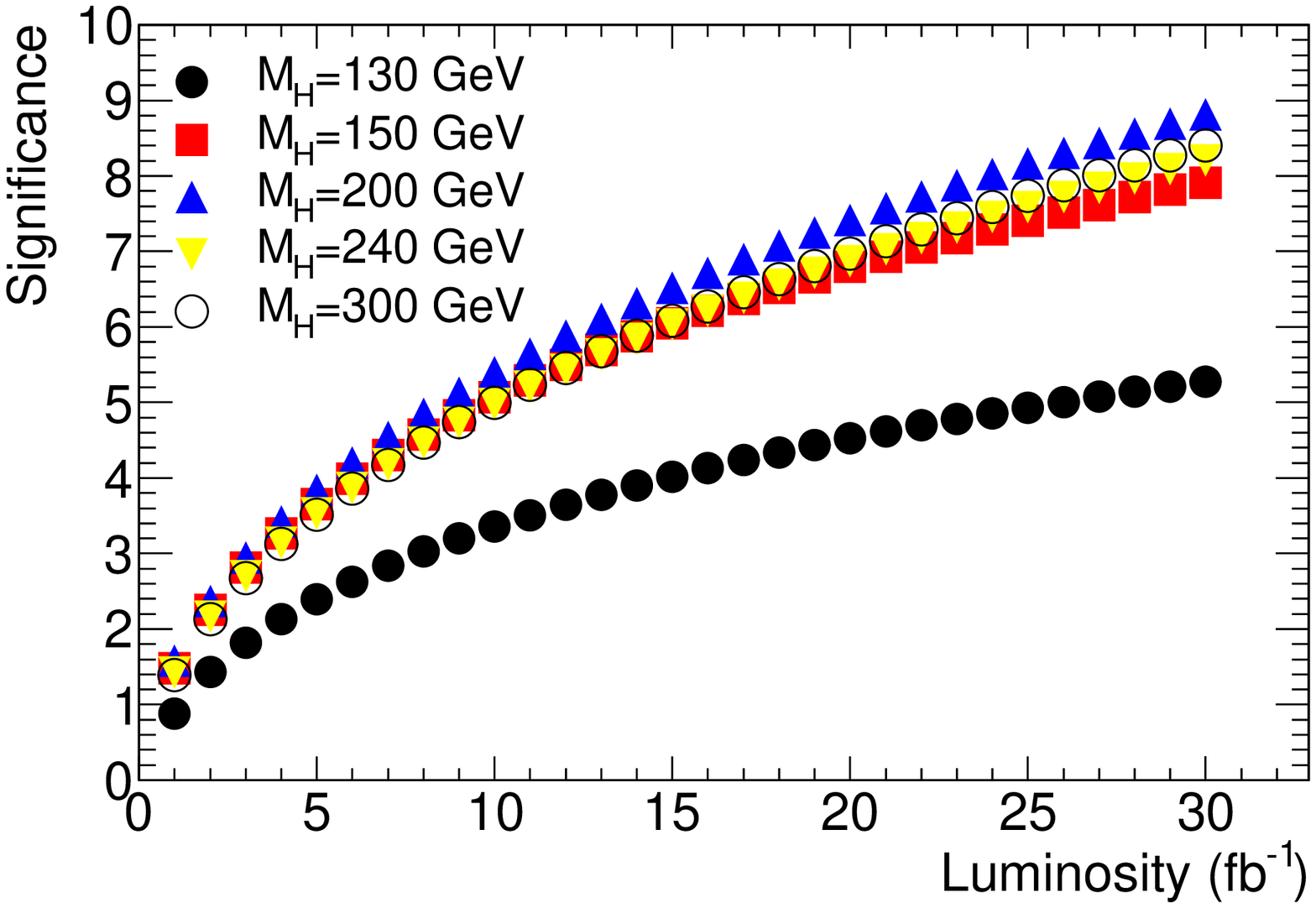}
}
\caption{
Standard Model Higgs significance as a function of its mass and as 
a function of the integrated luminosity for a single LHC experiment.}
\label{signif}
\end{figure}

The SM Higgs discovery potential depends on 
the luminosity and the mass of the Higgs. 
Here we assume a 7~TeV center of mass energy and an 
integrated luminosity, ranging from 1 to 30~fb$^{-1}$.
The significance can be calculated assuming the expected Higgs 
invariant mass distribution including detector resolution and 
the natural width of the SM Higgs.
The obtained significance as a function of mass and as 
a function of luminosity is presented in Figure~\ref{signif} 
for a single LHC experiment. The low Higgs mass search is 
mostly sensitive at 150~GeV where a 3$\sigma$ evidence should 
be found with 3~fb$^{-1}$ of integrated luminosity.
The real strength of this channel is in the high mass search.
Here one LHC experiment will need 2~fb$^{-1}$ in order to 
observe a 3$\sigma$ evidence for Higgs masses up to 300 GeV.
This is an important result since our method is data driven 
trying to maximize the information provided by the data.

In the remainder of this section we summarize the uncertainties 
of the method. At low luminosities below 5 fb$^{-1}$,
the main uncertainty is due to the limited data 
statistics in the M$_{34}$ and M$_{4l}$ plots. 
The expectation for the individual background components is as follows:
\begin{itemize}
\item
ZZ: for low luminosities, we expect only a handful of ZZ events for 
which we will have an excellent understanding of the M$_{34}$ shape 
thanks to existing Z$\rightarrow$2$\ell$ data. This leads to an extracted 
normalization for ZZ and ZZ* that is statistically dominated.
In addition, the ZZ* part is sensitive to theoretical and experimental 
uncertainties (such as control of lepton efficiencies).
\item
ZQj and Zjj backgrounds: the uncertainties are smaller due 
to the high statistics in the data control regions. In this case the 
dominant uncertainties are coming from the control of the 
discriminating variables used to go from the data control regions 
to the Higgs signal region. These variables include shower shapes, 
lepton impact parameter and tracking quality. 
Since these backgrounds are subdominant (ZQQ and ZZ$^{(*)}$ dominate), 
their impact to the total systematic uncertainty is 
expected to be small.
\item
ZQQ background: it is significant in the low mass range. The 
2D fit can constrain both the shape of ZQQ and its normalization.
Thus the ZQQ normalization and shape uncertainties are 
also dominated by the statistics in M$_{34}$ and M$_{4l}$, 
as well as by the systematic uncertainties on 
the other backgrounds. 
\end{itemize}

\section{Summary and Conclusions}
In this paper, a data-driven method for extracting a potential 
Higgs signal and its dominant
backgrounds in the search for Higgs boson decays to four leptons, 
was presented. 
The method relies on a combined unbinned likelihood fit of 
the 2-lepton, Z$^{(*)}$, and 4-lepton invariant masses.
The unique feature of the method is the attempt to 
maximize the use of the
information provided in the data: 
the 2D fit is assisted by normalization of the main Z+X backgrounds 
in control regions. 

The Zjj and ZjQ components are predicted using 
data control regions defined by a subset of the analysis event selection. 
The 2D fit proceeds assuming these normalizations of Zjj and ZjQ and 
their uncertainties, leaving the ZQQ normalization 
free. The ZZ$^{(*)}$ background is normalized by the ZZ peak in M$_{34}$ 
with an uncertainty dominated by the statistics in the observed 
diboson peak. The ZZ$^{(*)}$ M$_{34}$ shape is taken from Monte Carlo 
and includes all relevant uncertainties: the experimental coming 
from lepton efficiency, e/$\mu$ energy scale etc, and the 
theoretical. For ZZ$^{(*)}$  
the dominant systematic uncertainty for the early LHC running 
is due to the ZZ$^{(*)}$ statistics. 
The method exploits the fact that the Higgs distribution makes 
a narrow peak in the M$_{4\ell}$ distribution, while the backgrounds 
are smoothly distributed either in M$_{4\ell}$ or in M$_{34}$. 
Hence the good control of the background normalization is the key 
in the search for the Higgs.
It should be noted that, given high enough statistics in the 4-lepton
invariant mass sideband, a 1D fit could be practically equally powerful, 
especially in the high mass Higgs search case.

The method was used to obtain the
Higgs discovery potential for the next few years of LHC running.
It can be immediately used in the Higgs search currently 
performed by the ATLAS and CMS experiments.

\end{document}